\documentclass[conference]{IEEEtran}
\usepackage{amsmath}
\usepackage{latexsym}
\usepackage{amssymb}

\title{Construction of wiretap codes from ordinary channel codes}
\author{%
\IEEEauthorblockN{Masahito Hayashi}
\IEEEauthorblockA{Graduate School of Information Sciences, Tohoku Univ., Japan\\
CQT, National Univ. of Singapore, Singapore\\
Email: hayashi@math.is.tohoku.ac.jp}
\and
\IEEEauthorblockN{Ryutaroh Matsumoto}
\IEEEauthorblockA{Dept.\ of Communications and Integrated Systems,\\
Tokyo Institute of Technology, 152-8550 Japan\\
Email: ryutaroh@rmatsumoto.org}}

\newtheorem{lemma}{Lemma}

\newtheorem{proposition}[lemma]{Proposition}
\newtheorem{definition}[lemma]{Definition}
\newtheorem{theorem}[lemma]{Theorem}

\newtheorem{remark}[lemma]{Remark}

\begin{document}
\maketitle
\begin{abstract} From an arbitrary given
channel code over a discrete or Gaussian memoryless channel,
we construct a wiretap code with the strong security.
Our construction can achieve the wiretap capacity under mild
assumptions.
The key tool is the new privacy amplification theorem
bounding the eavesdropped information in terms of
the Gallager function.
\end{abstract}

\section{Introduction}
The information theoretical security \cite{liang09}
recently has attracted
huge interest.
The wiretap channel \cite{csiszar78,wyner75} is one of
its fundamental problems.
On a wiretap channel,
signals from the legitimate sender, called Alice,
is delivered to both legitimate receiver, called Bob, and
eavesdropper, called Eve.
The goal of Alice is to deliver messages to Bob
with low decoding probability
while keeping Eve from knowing much about the messages.
The capacity of wiretap channels has been determined for
discrete memoryless channels \cite{csiszar78,wyner75}
and for Gaussian channels \cite{lyc78} with a weaker notion of
security. The capacity of the above wiretap channels are also
determined with a stronger notion of security \cite{barros08,csiszar96,hayashi06b}.
The exponential decreasing rate of eavesdropped information is
also evaluated in \cite{hayashi06b,hayashi09b}.
Shannon theoretic study of the wiretap channels is fairly advanced.

On the other hand, there is still room for research in
the actual construction of codes for the wiretap channels,
which we call the wiretap codes.
Thangaraj et~al.\ \cite{thangaraj07}
proposed an LDPC based construction for specific
discrete memoryless channels,
and Klinc et~al.\ \cite{klinc09} proposed another LDPC based
construction for Gaussian channels.
Hamada \cite{hamada09} and Hayashi \cite{hayashi09b}
proposed general linear code based construction for additive
discrete memoryless channels.
Muramatsu and Miyake proposed a construction based
on the hashing property of LDPC matrices \cite{muramatsu09},
whose decoding requires the high-complexity minimum entropy decoder.

In those constructions except \cite{hayashi09b},
error correction and provision of secrecy are
combined in the constructed coding scheme.
This prevents us from using well-studied error-correcting codes
for the error correction in the wiretap codes, and
we need to adjust existing error-correcting codes or invent
a new wiretap code. This inconvenience may not be necessary.
In fact, in the quantum key distribution protocols,
the error correction and the provision of secrecy can be
separately studied and developed, see \cite{matsumoto09ldpc}
and references therein.

Moreover, previous constructions for discrete memoryless channels
do not cover all the discrete memoryless channels except \cite{muramatsu09}.
It is desirable
to have a construction of wiretap codes that can be used for
any discrete memoryless channels.

In this paper, we show two constructions of wiretap codes
from encoder and decoder in an ordinary channel code.
We do not modify the channel encoder nor decoder.
We attach the two-universal hash function to the encoder and
the decoder in order to realize secrecy from Eve.
We show that our construction can achieve the wiretap capacity
in the \emph{strong} security sense over discrete and Gaussian memoryless
channels,
while some of previous constructions do not have proofs of
the strong security.

The key tools for our constructions are the new forms
of the privacy amplification (PA) theorem \cite{bennett95privacy}.
The original PA theorem \cite{bennett95privacy}
does not achieve the optimal rate of PA,
which is the conditional Shannon entropy of Alice's  information
conditioned on Eve's information. Renner \cite{rennerphd}
improved it so that Renner's version of the theorem
can achieve the optimal rate. However, it does not enable us to
construct the wiretap code using an existing channel code.
The reason is that we cannot numerically compute the necessay
rate of hashing for a given channel code in order for
Eve's information on secret message to become sufficiently
small.
So we present two new forms of the PA theorem.
One is already given in \cite{hayashi09b}.
However, it requires the random selection of a chennel encoder
from the given family of channel codes.
We shall provide another form of the PA theorem in Theorem \ref{newresolvability},
which enables us to construct a wiretap code from single channel encoder.
Our new PA theorem is a nontrivial adaptation of
the channel resolvability lemma \cite[Lemma 2]{hayashi06b}.

This paper is organized as follows:
In Sec.\ \ref{sec2} we fix notations used in this paper.
In Secs.\ \ref{sec3} and \ref{sec4} two constructions of
wiretap codes are given.
In Sec.\ \ref{sec51} we present a novel privacy amplification theorem
bounding the eavesdropped information in terms of the Gallager
function.
Section \ref{sec5} concludes the paper.

\section{Preliminary}\label{sec2}
In this section we shall fix notations used in this paper
and review necessary prior results.
Let $\mathcal{X}$ be the finite alphabet of channel inputs,
$\mathcal{Y}$ the alphabet of channel outputs to
the legitimate receiver, called Bob,
and $\mathcal{Z}$ the alphabet of channel outputs to
the eavesdropper, called Eve.
The legitimate sender is called Alice.
We fix the conditional probability or conditional probability
density $Q_{Y|X}$ of the channel to Bob
and $Q_{Z|X}$ of the channel to Eve.
We assume channels are memoryless and further assume that
\begin{itemize}
\item both $\mathcal{Y}$ and $\mathcal{Z}$ are finite,
which means that the channels are discrete memoryless,
\item or $\mathcal{Y} = \mathcal{Z} = \mathbf{R}$ and
the channels are additive Gaussian.
\end{itemize}
Let $\mathcal{M}_n$ be the set of messages transmitted to Bob
secretly from Eve, $\eta_{\mathrm{Alice},n}$ a stochastic map
from $\mathcal{M}_n$ to $\mathcal{X}^n$ of a wiretap encoder,
and $\eta_{\mathrm{Bob},n}$ a deterministic map from
$\mathcal{Y}^n$ to $\mathcal{M}_n$.
We use the natural logarithm instead of $\log_2$ for convenience.

\begin{definition}
A rate $R >0$ is said to be achievable if
there exists a sequence $(\eta_{\mathrm{Alice},n}$,
$\eta_{\mathrm{Bob},n})$ of encoders and decoders such that
\begin{align*}
&\lim_{n\rightarrow \infty} \mathrm{Pr}[M_n \neq
\eta_{\mathrm{Bob},n}(\eta_{\mathrm{Alice},n}(M_n))] = 0,\\
&\lim_{n\rightarrow \infty} I(M_n; Z_n) = 0,\quad \liminf_{n\rightarrow \infty} \ln |\mathcal{M}_n| \geq R,
\end{align*}
where $M_n$ is the uniform random variable over $\mathcal{M}_n$
and $Z_n$ is the random variable for Eve's channel output
from channel input $\eta_{\mathrm{Alice},n}(M_n)$.
The supremum of the achievable rates is the capacity of
the wiretap channel $(Q_{Y|X}$, $Q_{Z|X})$.
\end{definition}
Note that we employ the strong security criterion
introduced by Csisz\'ar \cite{csiszar96} and
Maurer and Wolf \cite{maurer00}.
The necessity for the strong security is given in \cite{barros08,maurer00}.

\begin{proposition} \cite{barros08,csiszar96,hayashi06b}
The capacity of
the wiretap channel $(Q_{Y|X}$, $Q_{Z|X})$
is
\begin{equation}
\max_{P_{T}, P_{X|T}} [I(T;Y) - I(T;Z)]. \label{wiretap-capacity}
\end{equation}
\end{proposition}
In the next section,
we shall show a construction of wiretap encoder
and decoder from arbitrary given channel encoder and
decoder. In the construction,
we assume that we are given $Q_{X|T}$ achieving the maximum of
Eq.\ (\ref{wiretap-capacity}).
Note that when the wiretap channel is Gaussian, it is degraded
and we can take $T=X$ without losing the optimality.
In the construction, we shall also use a family of the two-universal
hash functions \cite{carter79}, which is reviewed next.

\begin{definition}
Let $\mathcal{S}_1$ and $\mathcal{S}_2$ be finite sets
and $\mathcal{F}$ a subset of the set of all mappings from
$\mathcal{S}_1$ to $\mathcal{S}_2$.
The family $\mathcal{F}$ is said to be a
family of two-universal hash functions if
\[
\mathrm{Pr}[F(x_1) = F(x_2)] \leq 1/|\mathcal{S}_2|,
\]
for all distinct $x_1$ and $x_2$ in $\mathcal{S}_1$,
where $F$ is the uniform random variable on $\mathcal{F}$.
\end{definition}

\section{Randomized construction of a wiretap code}\label{sec3}
\subsection{Encoder and decoder}
In this section we shall construct wiretap encoder and
decoder from arbitrary given ordinary channel encoder and decoder.
The construction in this section can achieve the wiretap capacity
(\ref{wiretap-capacity}) if the uniform distribution on $\mathcal{T}$ realizes
the wiretap capacity (\ref{wiretap-capacity}).
The assumptions are:
\begin{itemize}
\item We know $Q_{X|T}$ achieving the maximum of
Eq.\ (\ref{wiretap-capacity}). Denote by $\mathcal{T}$ the alphabet of $T$.
\item We are given a family channel encoders $\mu_{\mathrm{Alice},n,g}$
indexed by $g \in \mathcal{G}_n$
mapping a message in the message set $\mathcal{L}_n$ to a codeword
in $\mathcal{T}^n$ and a channel decoder $\mu_{\mathrm{Bob},n,g}$
mapping a received signal in $\mathcal{Y}^n$ to a message
in $\mathcal{L}_n$. The channel encoder $\mu_{\mathrm{Alice},n,g}$ is
a one-to-one map, and $\mathcal{T}^n$ is equal to
the disjoint union of $\mu_{\mathrm{Alice},n,g}(\mathcal{L}_n)$ for
$g \in \mathcal{G}_n$.
\item We are given a family $\mathcal{F}_n$ of two-universal
hash functions from $\mathcal{L}_n$ to $\mathcal{M}_n$,
where $\mathcal{M}_n$ is the message set of the wiretap code.
\end{itemize}

\begin{remark}
The assumption on the channel encoders is usually met
with linear codes.
We usually use the codebook of a linear code
whose codewords have zero syndrome.
If we allow codebooks to have nonzero syndrome,
then the family of codebooks with multiple syndromes
constitutes the family of encoders
$\{ \mu_{\mathrm{Alice},n,g} \mid g \in \mathcal{G}_n \}$.
\end{remark}

{From} these assumptions,
we can construct a wiretap encoder, which is
an extension of Hayashi's construction \cite{hayashi09b}.
Choose a hash function $F_n$ uniformly randomly from $\mathcal{F}_n$
and $G \in \mathcal{G}_n$.
For a given message $M_n$ to the wiretap encoder of code length $n$,
choose a message $L_n$ 
uniformly randomly from $F_n^{-1}(M_n) \subset \mathcal{L}_n$,
and compute the codeword $T_n=\mu_{\mathrm{Alice},n,G}(L_n)$ from
the channel encoder.
Finally, compute the actually transmitted signal $X_n$
by passing $T_n$ to the artificial memoryless channel $Q_{X|T}^n$.
The decoder maps a given received signal $Y_n$ in $\mathcal{Y}^n$
to the message $F_n(\mu_{\mathrm{Bob},n}(Y_n)) \in \mathcal{M}_n$.

The random selection of $F_n$ and $G_n$ is a fatal problem
because it requires sharing of common randomness
between Alice and Bob. However, we shall show that
$I(M_n;Z_n|F_n,G_n)$ can be upper bounded by an arbitrary positive
number $\epsilon_1 \times \epsilon_2$,
which means that at least $100(1-\epsilon_1)$\% choices of $f_n \in \mathcal{F}_n$ and $g_n \in \mathcal{G}_n$
keep $I(M_n;Z_n|F_n=f_n$, $G_n=g_n)$ below $\epsilon_2$.
Thus the legitimate sender and receiver can agree on the random choice of
$f_n$ before transmission of the secret messsage $M_n$.

\subsection{Evaluation of the eavesdropped information}
It should be clear that the (block) average decoding error probability
of the constructed wiretap code is lower than or equal to that of the
underlying code $(\mu_{\mathrm{Alice},n,g}$, $\mu_{\mathrm{Alice},n,g})$
for $g\in \mathcal{G}_n$
regardless of random choices of $F_n$ and $L_n$ from $M_n$.
The remaining task is evaluation of the eavesdropped information
$I(M_n, Z_n)$, where $Z_n$ is Eve's received signal
on the channel input $X_n$. To do so,
we introduce Hayashi's version of the privacy amplification
theorem \cite{hayashi09b}

\begin{proposition}\label{thm:hayashi09b}
Let $L$ be the \emph{uniform} random variable with a finite alphabet
$\mathcal{L}$ and $Z$ any random variable. If $Z$ is not discrete random variable
then the conditional probability of $Z$ given $L$ is assumed to be Gaussian.
Let $\mathcal{F}$ be a family of two-universal
hash functions from $\mathcal{L}$ to $\mathcal{M}$,
and $F$ be the uniform random variable on $\mathcal{F}$.
Then
\[
H(F(L)|F,Z)  \geq  \ln |\mathcal{M}| - 
\frac{|\mathcal{M}|^s\times \exp(\psi(s,P_{LZ}))}{s|\mathcal{L}|^s}
\]
for $0<s\leq 1$, where
\[
\psi(s,P_{LZ}) = \ln \sum_{z}\frac{\sum_{\ell} P_L(\ell)(P_{Z|L}(z|\ell))^{1+s}}{P_Z(z)^s}.
\]
If $Z$ is conditionally Gaussian $\sum_z$ should be replaced by the integration
and $P_Z$, $P_{Z|L}$ denote probability densities.
\end{proposition}

\begin{remark}
The above proposition is a combination of \cite[Eq.\ (2)]{hayashi09b}
and the argument in proof of \cite[Theorem 2]{hayashi09b}.
It was assumed that $Z$ was discrete in \cite{hayashi09b}.
However, when the conditional probability of $Z$ given $L$ is
Gaussian, there is no difficulty to extend the original result.
It should be also noted that the uniformity assumption 
on $L$ is indispensable, otherwise the claim is false.
\end{remark}

By the above proposition, for fixed $G=g\in \mathcal{G}_n$
we have
\begin{eqnarray}
I(M_n;Z_n^g,F_n)&=&I(M_n;Z_n^g|F_n)\nonumber\\
 &=& H(M_n|F_n) - H(M_n | Z_n^g,F_n)\nonumber\\
&\leq& \ln |\mathcal{M}_n| - H(M_n | Z_n^g,F_n) \nonumber\\
& \leq &
\frac{|\mathcal{M}_n|^s\times \exp(\psi(s,P^g_{L_nZ_n}))}{|\mathcal{L}_n|^ss}
\label{eq:diff}
\end{eqnarray}
for $0 < s \leq 1$, where
$P^g_{L_nZ_n}$ is the joint probability distribution
and $Z_n^g$ is Eve's received signal
with a fixed $g \in \mathcal{G}_n$

A major problem with the last upper bound (\ref{eq:diff}) on 
$I(M_n;Z_n|F_n)$ is that for a given channel code it is
practically impossible to numerically compute
$\psi(s,P^g_{L_nZ_n})$. To overcome this difficulty
we shall upper bound $\exp(\psi(s,P^g_{L_nZ_n}))$ by 
$\exp(\psi(s,P_{TZ}))$, where $P_{TZ}$ is a joint distribution
on $\mathcal{T}\times \mathcal{Z}$.

Let $T_g = \mu_{\mathrm{Alice},n,g}(L_n)$ that is a random variable
on $\mathcal{T}^n$. Note that
$T_g$ is the uniform random variable
on $\mu_{\mathrm{Alice},n,g}(\mathcal{L}_n) \subset \mathcal{T}^n$.
By the assumption on the given family of
channel encoders $\mu_{\mathrm{Alice},n,g}$, $g \in \mathcal{G}_n$,
the convex combination of
$\sum_{g\in\mathcal{G}_n} P_{T_g}/|\mathcal{G}_n|
$
is the uniform distribution $\mathrm{Uniform}(\mathcal{T}^n)$
on $\mathcal{T}^n$.
By the concavity of $\exp(\psi(s,\cdot))$ on the channel input probability
distribution\footnote{The concavity is proved under that
assumption that $\mathcal{Z}$ is finite. However,
if the conditional probability $Q_{Z|X}$ is Gaussian,
the concavity proof needs no change except notational ones.} \cite[Lemma 1]{hayashi09b}, we have
\begin{align*}
\frac{1}{|\mathcal{G}_n|}\sum_{g\in\mathcal{G}_n}\exp(\psi(s,P^g_{L_nZ_n}))
\leq& \exp(\psi(s,Q_{Z|T}^n\mathrm{Uniform}(\mathcal{T}^n))\\
=&  \exp(n\psi(s,Q_{Z|T}\mathrm{Uniform}(\mathcal{T})).
\end{align*}
Observe that computation of the last mathematical expression is easy
for almost all channels.

What we have proved is
\begin{equation}
I(M_n;Z_n|F_n,G_n)
\leq 
\frac{|\mathcal{M}_n|^s\times \exp(n\psi(s,Q_{Z|T}\mathrm{Uniform}(\mathcal{T})))}{|\mathcal{L}_n|^ss}. \label{upper1}
\end{equation}
Observe that the minimization of the RHS of Eq.\ (\ref{upper1})
over $s$ is also computable by the bisection method \cite[Algorithm 4.1]{boyd04} because it is convex with respect to $s$.
The logarithm of the right hand side is
\begin{equation}
s \left(\ln |\mathcal{M}_n|-\ln |\mathcal{L}_n| + \frac{n\psi(s,Q_{Z|T}\mathrm{Uniform}(\mathcal{T}))}{s}\right) - \ln s. \label{eq3}
\end{equation}
By l'H\^opital's theorem,
we have
\[
\lim_{s\rightarrow +0}\frac{\psi(s,Q_{Z|T}\mathrm{Uniform}(\mathcal{T}))}{s} = 
I(\mathrm{Uniform}(\mathcal{T}),Q_{Z|T}),
\]
where the right hand side is the mutual information between the channel
output and the uniform channel input to the imaginary channel $Q_{Z|T}$.
Thus,  by  choosing $s$ such that
$\frac{\psi(s,Q_{Z|T}\mathrm{Uniform}(\mathcal{T}))}{s} <
I(\mathrm{Uniform}(\mathcal{T}),Q_{Z|T}) + \delta$,
we can see that
if $\ln |\mathcal{M}_n|  < \ln|\mathcal{L}_n -n (I(\mathrm{Uniform}(\mathcal{T}),Q_{Z|T})+\delta)$ for some $\delta > 0$
then
Eq.~(\ref{eq3}) converges to $-\infty$ as $n\rightarrow \infty$,
which means the eavesdropper Eve has little information on the
secret message. This means that
if $\ln |\mathcal{L}_n|/n$ converges to $I(\mathrm{Uniform}(\mathcal{T}),Q_{Z|T})$ and the wiretap capacity (\ref{wiretap-capacity}) is achieved with
uniform channel input then this construction
also achieves the wiretap capacity.

Drawbacks in the proposed construction is the random selection
of channel encoders.
This requires that almost all pairs of encoder and decoder
have to provide low decoding error probability,
which is not verified with most of channel codes.
Moreover, in some case, for example the channel encoder
using the Trellis shaper \cite{forney92},
it is difficult to prepare a family of encoders that
satisfies the requirement.
Thus, in the next section, we show a deterministic construction
of a wiretap code from a given channel code.

\section{Deterministic construction of a wiretap code}\label{sec4}
In this section,
we assume that the index set $\mathcal{G}_n$ has only one element,
and we are given a pair of an encoder $\mu_{\mathrm{Alice},n}$
a decoder $\mu_{\mathrm{Bob},n}$.
We also assume that the given family $\mathcal{F}_n$ of hash functions
satisfies the condition that for all $f \in \mathcal{F}_n$ and $m \in \mathcal{M}_n$
we have $|f^{-1}(m)| = |\mathcal{L}_n|/|\mathcal{M}_n|$ in order to
apply Theorem \ref{newresolvability} in Sec.\ \ref{sec51}.
This assumption on $\mathcal{F}_n$ is satisfied, for example when
$\mathcal{M}_n = \mathbf{F}_q^k$ and $\mathcal{L}_n = \mathbf{F}_q^n$,
using the set of all the surjective linear maps from $\mathcal{L}_n$
to $\mathcal{M}_n$. Moreover, the linear mappings defined by
the concatenation of the identity matrix and the Toeplitz matrix
considered in \cite[Appendix]{hayashi09b} also satisfy the assumption
and is more efficiently implemented in practice.

The construction of the wiretap code is the same as the
previous section except that there is no random selection
of encoders.
The construction in this section can achieve the
wiretap capacity (\ref{wiretap-capacity})
if the distribution $P_T$ on $\mathcal{T}$ realizing
(\ref{wiretap-capacity}) also maximizes
the mutual information $I(P_T, Q_{Z|T})$ to the eavesdropper.
In order to evaluate the average of the mutual information, we develop 
a new privacy amplification theorem (Theorem \ref{newresolvability}) based on Gallager function by modifying \cite[Lemma 2]{hayashi06b}
in the next section.
Applying this result, 
one can show that
\[
I(M_n;Z_n|F_n)
\leq
\frac{|\mathcal{M}|^s \exp(\phi(s,Q^n_{Z|T},P_{T_n}))}{|\mathcal{L}|^ss},
\]
for $0 \leq s \leq 1/2$, where
\begin{eqnarray*}
&&\phi(s,Q^n_{Z|T},P_{T_n}) \\
&=& \ln \int_{\mathcal{Z}^n}\left(
\sum_{t\in\mathcal{T}^n} P_{T_n}(t) (Q^n_{Z|T}(z|t))^{1/(1-s)}\right)^{1-s}dz.
\end{eqnarray*}
If $\mathcal{Z}$ is finite, the integration should be replaced by
summation and $Q_{Z|T}$ should be interpreted as the conditional
probability.

Again, for a given channel encoder $\mu_{\mathrm{Alice},n}$,
it is also practically impossible to compute
$\phi(s,Q^n_{Z|T},P_{T_n})$.
We shall show that a method to upper bound it.
We have
\[
\exp(\phi(s,Q^n_{Z|T},P_{T_n}))
\leq
\max_{P_n} \exp(\phi(s,Q^n_{Z|T},P_n)),
\]
where $P_{n}$ is a probability distribution on $\mathcal{T}_n$.
Observe that $\phi$ is essentially same as the
function $E_0$ in \cite{arimoto73,gallager68}.
Thus if $P_{1,s}$ maximizes $\exp(\phi(s,Q_{Z|T},P_{1,s}))$,
then its $n$-fold i.i.d. extension $P_{1,s}^n$ also maximizes
$\max_{P_n} \exp(\phi(s,Q^n_{Z|T},P_n))$ \cite{arimoto73},
and we have
\begin{equation}
I(M_n;Z_n|F_n)
\leq
\frac{|\mathcal{M}|^s \exp(n\phi(s,Q_{Z|T},P_{1,s}))}{|\mathcal{L}|^ss}.
\label{eq10}
\end{equation}
Observe that for fixed $s$ and $Q_{Z|T}$, $\exp(\phi(s,Q_{Z|T},P_{1,s}))$ 
is a concave function on a convex set and $P_{1,s}$ can easily be computed
\cite{boyd04}. Observe also that
for fixed $Q_{Z|T}$, the function $\max_{P_{1,s}}$[RHS of Eq.\ (\ref{eq10})]
is a convex function of $s$, thus $\min_s\max_{P_{1,s}}$[RHS of Eq.\ (\ref{eq10})]
can also be easily computed by the bisection method \cite[Algorithm 4.1]{boyd04}.

The logarithm of the right hand side is
\[
s \left(\ln |\mathcal{M}_n|-\ln |\mathcal{L}_n| + \frac{n\phi(s,Q_{Z|T},P_{1,s})}{s}\right) - \ln s. 
\]
Since $\phi$ is essentially $E_0$ in \cite{gallager68},
$\lim_{s\rightarrow 0} \phi(s,Q_{Z|T},P)/s = I(P, Q_{Z|T})$, where
$P$ is a distribution on $\mathcal{T}$.
Let $P_\mathrm{max}$ be a distribution on $\mathcal{T}$ maximizing
$I(P, Q_{Z|T})$.
Therefore, by the almost same argument as Section \ref{sec2},
if $\ln|\mathcal{M}_n| < \ln |\mathcal{L}_n| - n (I(P_\mathrm{max}, Q_{Z|T})+\delta)$
for all $n$, then $I(M_n;Z_n|F_n)$ goes to zero as $n\rightarrow \infty$.
If $P_\mathrm{max}$ also maximizes the wiretap capacity (\ref{wiretap-capacity})
and the given channel code achieves the information rate
$I(P_\mathrm{max}, Q_{Y|T})$ then the construction in this section achieves the
wiretap capacity.

\def\rE{{\rm E}}

\allowdisplaybreaks[1]

\section{New privacy amplification theorem in terms of the Gallager function}\label{sec51}
We shall show the following new privacy amplification theorem
that is indispensable with the deterministic construction of wiretap
codes in Sec.\ \ref{sec4}.
\begin{theorem}\label{newresolvability}
Assume that the given family of two-universal hash function $F$ from $\mathcal{L}$
to $\mathcal{M}$ satisfies that
\begin{eqnarray*}
|F^{-1}(m)|=\frac{|\mathcal{L}|}{|\mathcal{M}|}, \quad
\forall m,
\end{eqnarray*}
a fixed conditional probability $Q_{Z|L}$ is given,
and
the random variable $L$ obeys the uniform distribution on $\mathcal{L}$.
Then,
\begin{align}
I(F(L); Z|F) = \rE_{F} I(F(L);Z)\le 
\frac{|\mathcal{M}|^s \exp(\bar{\phi}(s,Q_{Z|L}))}{|\mathcal{L}|^s s},
\label{haya-1}
\end{align}
for $0 \leq s \leq 1/2$, where
$\rE_F$ expresses the expectation concerning the random variable $F$,
\[
\bar{\phi}(s,Q_{Z|L}) 
= \ln \int_{\mathcal{Z}}\Bigl(
\rE_L (Q_{Z|L}(z|L))^{1/(1-s)}\Bigr)^{1-s}
d z
\]
and $dz$ is an arbitrary measure.
\end{theorem}

\noindent\emph{Proof.}
Observe first that the joint probability $P_{FL}=P_F \times P_L$ and
the conditional probability $Q_{Z|L}$ uniquely determines
$Q_{Z|F(L)}$.
We can check that 
the function $s \mapsto \bar{\phi}(s,Q^n_{Z|F(L)})$
satisfies the following properties:
\begin{align*}
& \bar{\phi}(0,Q_{Z|F(L)})= 0 , \quad
\frac{d^2 \bar{\phi}(s,Q_{Z|F(L)})}{d s^2}
 \ge 0 \\
& \frac{d \bar{\phi}(s,Q_{Z|F(L)})}{d s}
\Bigr|_{s=0} = I(F(L);Z).
\end{align*}
Hence, its convexity guarantees
the inequality
$s \rE_{F}I(F(L);Z) \le \rE_{F} \bar{\phi}(s,Q_{Z|F(L)})$,
which implies the inequality
\begin{align}
\rE_{F}I(F(L);Z)
\le  \rE_{F} \frac{\bar{\phi}(s,Q_{Z|F(L)})}{s}\label{haya-2}
\end{align}
for $0 < s \leq \frac{1}{2}$.
In the following, we denote the uniform distriburtion on $\mathcal{L}$ by $P_{L}$

Let $1+u= \frac{1}{1-s}$, then 
$1\ge u > 0$ and $s= \frac{u}{1+u}$.
Since $x \mapsto x^u$ is concave,
\begin{align}
& \rE_F \bigr[ \sum_{\ell':F(\ell')=F(\ell), \ell'\neq \ell } Q_{Z|L} (z|\ell')\bigr]^u 
\nonumber \\
\le &\bigl[\rE_F \sum_{\ell':F(\ell')=F(\ell), \ell'\neq \ell } Q_{Z|L} (z|\ell') \bigl]^u \nonumber \\
\le & \bigl[\sum_{\ell': \ell'\neq \ell } \frac{1}{|\mathcal{M}|}Q_{Z|L} (z|\ell') \bigl]^u
\le \bigl[\frac{|\mathcal{L}|}{|\mathcal{M}|}Q_{Z} (z) \bigl]^u 
= (\frac{|\mathcal{L}|}{|\mathcal{M}|})^u Q_{Z} (z) ^u. 
\label{haya-3}
\end{align}
Using (\ref{haya-3}) and the relation $(x+y)^u \le x^u + y^u$ for 
two positive real numbers $x,y$, 
we obtain
\begin{align}
& e^{\rE_F \bar{\phi}(s,Q_{Z|F(L)})}
\le \rE_F e^{\bar{\phi}(s,Q_{Z|F(L)})} \label{haya-9}\\
=&  \rE_F \int_{\mathcal{Z}} 
\Bigl( \sum_{m\in\mathcal{M} } \frac{1}{|\mathcal{M}|} 
Q_{Z|F(L)}(z|m )^{1+u}\Bigr)^{\frac{1}{1+u}}
d z \nonumber \\
\le & 
\int_{\mathcal{Z}} 
\Bigl( \rE_F \sum_{m\in\mathcal{M} } \frac{1}{|\mathcal{M}|} 
Q_{Z|F(L)}(z|m )^{1+u}\Bigr)^{\frac{1}{1+u}}
d z \label{haya-8} \\
= & 
\int_{\mathcal{Z}} 
\Bigl( \rE_F \sum_{m\in\mathcal{M} } \frac{1}{|\mathcal{M}|} 
Q_{Z|F(L)}(z|m )
Q_{Z|F(L)}(z|m )^{u}
\Bigr)^{\frac{1}{1+u}}
d z \nonumber \\
= & 
\int_{\mathcal{Z}} 
\Bigl( \rE_F 
\sum_{m\in\mathcal{M} } \frac{1}{|\mathcal{M}|} 
\Bigl[
\sum_{\ell\in\mathcal{L}:F(\ell)=m } \frac{|\mathcal{M}|}{|\mathcal{L}|} 
Q_{Z|L}(z|\ell)
\Bigr]  \nonumber
\\ & \qquad  \Bigl[
\sum_{\ell\in\mathcal{L}:F(\ell)=m } \frac{|\mathcal{M}|}{|\mathcal{L}|} 
Q_{Z|L}(z|\ell)
\Bigr]^u
\Bigr)^{\frac{1}{1+u}}
d z \nonumber \\
= & 
\int_{\mathcal{Z}} 
\Bigl( \rE_F 
\sum_{\ell\in\mathcal{L} } \frac{1}{|\mathcal{L}|} 
Q_{Z|L}(z|\ell)
(\frac{|\mathcal{M}|}{|\mathcal{L}|})^u
\Bigl[
Q_{Z|L}(z|\ell)
 \nonumber \\ & \qquad  +
\sum_{\ell'\in\mathcal{L}:F(\ell')=F(\ell),\ell'\neq \ell } 
Q_{Z|L}(z|\ell')
\Bigr]^u
\Bigr)^{\frac{1}{1+u}}
d z \nonumber \\
\le & 
\int_{\mathcal{Z}} 
\Bigl( 
\rE_F 
\sum_{\ell\in\mathcal{L} } \frac{1}{|\mathcal{L}|} 
Q_{Z|L}(z|\ell)
(\frac{|\mathcal{M}|}{|\mathcal{L}|})^u
\Bigl[
Q_{Z|L}(z|\ell)^u  \nonumber \\ & \qquad 
+
\bigl(\sum_{\ell'\in\mathcal{L}:F(\ell')=F(\ell),\ell'\neq \ell } 
Q_{Z|L}(z|\ell')\bigr)^u \Bigr]
\Bigr)^{\frac{1}{1+u}}
d z \label{haya-5}\\
= & 
\int_{\mathcal{Z}} 
\Bigl( 
(\frac{|\mathcal{M}|}{|\mathcal{L}|})^u
\sum_{\ell\in\mathcal{L} } \frac{1}{|\mathcal{L}|} 
Q_{Z|L}(z|\ell)^{1+u} +
\bigl(\frac{|\mathcal{M}|}{|\mathcal{L}|}\bigr)^u  \nonumber \\ &  
\times \sum_{\ell\in\mathcal{L} } \frac{1}{|\mathcal{L}|} 
Q_{Z|L}(z|\ell)
\rE_F \bigl( 
\sum_{\ell\neq \ell'\in F^{-1}(\ell) } 
Q_{Z|L}(z|\ell')
\bigr)^u 
\Bigr)^{\frac{1}{1+u}}
d z \nonumber \\
\le & 
\int_{\mathcal{Z}} 
\Bigl( 
(\frac{|\mathcal{M}|}{|\mathcal{L}|})^u
\rE_L Q_{Z|L}(z|L)^{1+u}  \nonumber \\ & \qquad 
+
(\frac{|\mathcal{M}|}{|\mathcal{L}|})^u
Q_{Z}(z)
\bigl(\frac{|\mathcal{L}|}{|\mathcal{M}|}\bigr)^u 
Q_{Z}(z)^u 
\Bigr)^{\frac{1}{1+u}}
d z \label{haya-4}\\
= & 
\int_{\mathcal{Z}} 
\Bigl( 
(\frac{|\mathcal{M}|}{|\mathcal{L}|})^u
\rE_L Q_{Z|L}(z|L)^{1+u}
+
Q_{Z}(z)^{1+u}
\Bigr)^{\frac{1}{1+u}}
d z \nonumber \\
\le & 
\int_{\mathcal{Z}} 
\Bigl( 
(\frac{|\mathcal{M}|}{|\mathcal{L}|})^u
\rE_L Q_{Z|L}(z|L)^{1+u}
\Bigr)^{\frac{1}{1+u}}
+
(Q_{Z}(z)^{1+u})^{\frac{1}{1+u}}
d z \label{haya-7} \\
= & 
\int_{\mathcal{Z}} 
(\frac{|\mathcal{M}|}{|\mathcal{L}|})^{\frac{u}{1+u}}
\Bigl( 
\rE_L Q_{Z|L}(z|L)^{1+u}
\Bigr)^{\frac{1}{1+u}}
+
Q_{Z}(z)
d z \nonumber \\
= & 
1+
(\frac{|\mathcal{M}|}{|\mathcal{L}|})^{\frac{u}{1+u}}
\int_{\mathcal{Z}} 
\Bigl( 
\rE_L Q_{Z|L}(z|L)^{1+u}
\Bigr)^{\frac{1}{1+u}}
d z \nonumber \\
= & 
1+
(\frac{|\mathcal{M}|}{|\mathcal{L}|})^{s}
e^{\bar{\phi}(s,Q^n_{Z|L})} \nonumber,
\end{align}
where the inequalities can be shown in the following way.
Ineq. (\ref{haya-4}) follows from (\ref{haya-3}).
Ineq. (\ref{haya-5}) and (\ref{haya-7}) follow from inequality $(x+y)^u \le x^u+y^u$ for $0 \le u \le 1$ and $x,y\ge 0$.
Ineq. (\ref{haya-8}) follows from the concavity of $x \mapsto x^u$ for $0 \le u \le 1$.
Ineq. (\ref{haya-9}) follows from the convexity of $x \mapsto e^x$.
Since the above inequality implies
\begin{align*}
\rE_F \bar{\phi}(s,Q_{Z|F(L)})
\le &
\ln
[1+
(\frac{|\mathcal{M}|}{|\mathcal{L}|})^{s}
e^{\bar{\phi}(s,Q^n_{Z|L})} ]\\
\le &
(\frac{|\mathcal{M}|}{|\mathcal{L}|})^{s}
e^{\bar{\phi}(s,Q^n_{Z|L})} ,
\end{align*}
using (\ref{haya-2}) we obtain (\ref{haya-1}).

\section{Conclusion}\label{sec5}
In this paper, starting from an arbitrary given channel code,
we showed two constructions of wiretap codes.
The first one involves the randomized selection of channel
encoders. The second one is deterministic.
These two construction can achieve the wiretap capacity
under different conditions.
Our constructions provide the strong security.

Ideally, the addition of hash functions to an arbitrary
given channel code should always achieve the wiretap capacity
whenever the given channel code achieves the capacity of
the composition of the artificially added channel $Q_{X|T}$ plus the physical
channel $Q_{Z|X}$. The proposed constructions fall short of this ideal.
The improved construction should be explored.
The numerical computation  of an optimal $Q_{X|T}$ from given
$Q_{Y|X}$ and $Q_{Z|X}$ is also an open problem.

\section*{Acknowledgment}
The second author would like to thank Prof.\ Yasutada Oohama,
Prof.\ Tomohiko Uyematsu,
Dr.\  Shun Watanabe, and Dr.\ Kenta Kasai for helpful discussions.
This research was partially supported by a Grant-in-Aid for Scientific Research
in the Priority Area ``Deepening and Expansion of Statistical Mechanical Informatics (DEX-SMI),'' No.\ 18079014
and a MEXT Grant-in-Aid for Young Scientists (A) No.\ 20686026.


\end{document}